\documentclass[prd,aps,preprint,showpacs,nofootinbib,superscriptaddress,tightenlines]{revtex4-1}
\usepackage{epsfig}
\usepackage{amssymb}
\usepackage{amsmath}
\usepackage{graphicx}
\usepackage{psfrag}

\begin{document}

\title{Initial and Final State Interaction Effects in Small-$x$ Quark
Distributions}

\author{Bo-Wen Xiao}
\affiliation{Nuclear Science Division, Lawrence Berkeley National
Laboratory, Berkeley, CA 94720}

\author{Feng Yuan}
\affiliation{Nuclear Science Division, Lawrence Berkeley National
Laboratory, Berkeley, CA 94720} \affiliation{RIKEN BNL Research
Center, Building 510A, Brookhaven National Laboratory, Upton, NY
11973}

\begin{abstract}
We study the initial and final state interaction effects in the
transverse momentum dependent parton distributions in the
small-$x$ saturation region. In particular, we discuss the quark
distributions in the semi-inclusive deep inelastic scattering,
Drell-Yan lepton pair production and dijet-correlation processes
in $pA$ collisions. We calculate the quark distributions in the
scalar-QED model and then extend to the color glass
condensate formalism in QCD. The quark distributions are found
universal between the DIS and Drell-Yan processes. On the other hand,
the quark distribution from the $qq'\to qq'$ channel contribution
to the dijet-correlation process is not universal. However, we find
that it can be related to the quark distribution in DIS process
by a convolution with the normalized unintegrated gluon distribution
in the color glass condensate formalism in the large $N_c$ limit.
\end{abstract}

\maketitle




\section{Introduction}

Partonic internal structure of nucleon and nucleus have attracted
many theoretical and experimental investigations in the past and
are still in the frontier of the subatomic physics research. These
studies aim at providing us accurate description of the hadronic
structure in terms of fundamental degree of freedom in Quantum
Chromodynamics (QCD), and meanwhile presenting an important path
to discover the new physics beyond the Standard Model, which are
currently undertaken at various high energy experiments, such as
the Fermilab Tevatron, and the Large Hadron Collider at CERN. One
of the most important objects is the parton distribution functions
(PDFs). These functions describe the internal structure of hadrons
in terms of the distribution of the longitudinal momentum fraction
$x$ carried by partons in the infinite momentum frame and the
relevant QCD factorization has been well
developed~\cite{ColSopSte89}.

In recent years, hadronic physics community have extended the Feynman parton
distributions to include the dependence on additional dimensions, in
particular in the transverse directions perpendicular to the parenting
hadron momentum direction. These extensions appear in two different
fashions: in the transverse coordinate space as the generalized parton
distributions (GPDs)~\cite{gpdreview}; in the transverse momentum space as the transverse
momentum dependent parton distributions (TMDs)~\cite{Ji:2003ak}. A number
of experimental facilities, such as the 12GeV upgrade of Jefferson Lab, the
Relativistic Heavy Ion Collider at the Brookhaven National Lab, and the planned
Electron-Ion Collider, are trying to measure these distribution functions.
These studies will lead us to the final answers to the important questions
concerning the nucleon (nucleus) structure: the proton spin and parton
saturation in nucleon (nucleus) at small $x$.

Transverse momentum dependence in the parton distributions is also crucial
to understand some novel hadronic physics phenomena in high energy
scattering processes. This includes, for example, the single transverse spin
asymmetries~\cite{BroHwaSch02,Col02,BelJiYua02,{Boer:2003cm}} and small-$x$
saturation phenomena~\cite{Brodsky:2002ue,{Iancu:2003xm}}. In the case for
the small-$x$ physics, the so-called $k_t$-dependent parton distributions
contain resummation effects which come from multiple scattering associated
with the nucleus target. Phenomenologically, the $k_t$-dependent gluon
distribution (also called unintegrated gluon distribution) function has been
applied to describe various high energy hadronic processes~\cite%
{Iancu:2003xm}.

In general, the initial and final state interaction effects associated with
the transverse momentum dependent parton distributions introduce additional
QCD dynamics in these processes. For example, they lead to the non-universality
for the transverse spin dependent TMD parton
distributions~\cite{BroHwaSch02,Col02,BelJiYua02,BoeVog03,
mulders,qvy-short,Collins:2007nk, Vogelsang:2007jk,{Rogers:2010dm}}.
In ref.~\cite{Xiao:2010sp}, we extended the universality discussions of TMD
parton distributions to the small-$x$ domain. In order to study the factorization issues
relevant to small-$x$ saturation physics, it is of advantage to focus on the
two scale processes, such as semi-inclusive hadron production in deep
inelastic scattering, Drell-Yan lepton pair production in hadronic
reactions, and the di-jet correlations in these processes. In particular, we
analyzed the small-$x$ transverse momentum dependent
quark distributions probed in hadronic dijet-correlation in nucleon-nucleus
collisions, as compared to that in the deep elastic lepton-nucleus (nucleon)
scattering. Due to the nuclear enhancement, any soft gluon exchanges
originated from the proton can be neglected. Thus, we only need to resum the
soft gluon exchanges with the nucleus target in the large nuclear number limit.
This procedure eventually helps us to
obtain an effective $k_{t}$ factorization with the modified nuclear parton
distributions in $pA$ collisions. There have been interesting experimental
results on di-hadron correlation in deuteron-gold collisions at RHIC,
where a strong back-to-back de-correlation was found in the forward
rapidity region of the deuteron as compared to the narrow back-to-back
peaks observed in the central rapidity region~\cite{rhic-data}.
The purposes of this paper are to derive in details the QED results which we
described in ref.~\cite{Xiao:2010sp} where we have shown the
non-universality of TMD parton distributions, and then generalize to small-$%
x $ models in QCD.

We will start with a scalar-QED model calculation.
There are a number of motivations for doing this. First,
QED model is calculable, which makes it straightforward to resum all order
initial and final state interaction effects in various processes. This
allows us to rigorously discuss these effects, and shed light on the real
QCD calculations. Second, there is similarity between the QED model
calculations and the QCD saturation models. Quite a few results in the former
framework can be directly generalized to the latter one.
Third, the QED calculations are important in their
own perspective. High order corrections in QED processes are interesting
topics and have attracted intensive investigations since the QED was founded
several decades ago. In particular, for the lepton pair production and
photon radiation processes associated with large nucleus, higher order QED
corrections have generated interests from both theory and experiment sides~\cite{qed-hic}.
Current running heavy ion collisions experiments are pursuing these studies
at both RHIC and LHC facilities. The theoretical investigations shall
provide further understanding of these processes.

In the saturation domain of QCD, the McLerran-Venugopalan model~\cite%
{McLerran:1993ni} describes high density small-$x$ partons in a relativistic
large nucleus by treating the nucleus as a set of randomly distributed color
sources $\rho _{a}\left( z^{-},z_{\perp }\right) $ which generate soft
classical gluon fields. Using the McLerran-Venugopalan model, we will demonstrate
that the $k_{t}$ dependent quark distributions at small-$x$ are not
universal  and the quark distributions involved in DIS
and di-jet production processes are distinct. Furthermore, we find a simple
formula to relate these two quark distributions through a convolution
with the normalized unintegrated gluon distribution.

The rest of the paper is organized as follows. In Sec. II, we construct a
scalar QED model to investigate the universality property for the small-$x$
parton distributions in various processes, where all gauge boson exchange
contributions can be summed up, including all initial and final state
interactions. In Sec. III, we demonstrate that the small-$x$ TMD quark distributions
of a large nucleus are not universal as well. In Sec.
IV, we summarize our results.

\section{Initial/final State Interactions in QED Models}

In this section, we discuss the universality issue for the transverse
momentum dependent parton distributions, by studying a scalar QED model. We
will first introduce the QED model for our calculations, and discuss the
universality issue of the TMD parton distribution functions in several processes.

\subsection{Model Description}

We follow Ref.~\cite{Brodsky:2002ue} to construct the model for our
calculations. This is a QED scalar model. It consists of heavy
$D$ and light $\phi $ charged scalars
with masses $M$ and $m$, respectively, interacting with massive U(1) gauge
fields $A_{\mu }$ with the mass $\lambda $,
\begin{equation}
\mathcal{L}_{\mathrm{sQED}}=\left( \mathcal{D}_{\mu }\phi _{1}\right)
^{\dagger }\mathcal{D}_{\mu }\phi _{1}+\left( \mathcal{D}_{\mu }\phi
_{2}\right) ^{\dagger }\mathcal{D}_{\mu }\phi _{2}+\left( \mathcal{D}_{\mu
}D\right) ^{\dagger }\mathcal{D}_{\mu }D-m^{2}\phi ^{\dagger }\phi
-M^{2}D^{\dagger }D-\frac{1}{4}F_{\mu \nu }^{2}+\frac{\lambda ^{2}}{2}A_{\mu
}^{2}\,,
\end{equation}
via the covariant derivative $\mathcal{D}_{\mu }\equiv \partial _{\mu
}+igA_{\mu }$. In the above equation, we introduced two charged scalar
particles, $\phi _{1}$ and $\phi _{2}$ with charges $g_{1}$ and $g_{2}$,
respectively. The purpose for this choice is to study the universality of
the parton distribution in dijet-correlation.

\begin{figure}[tbp]
\begin{center}
\includegraphics[width=12cm]{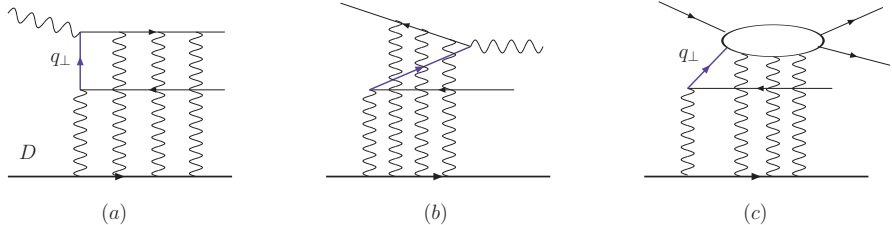}
\end{center}
\caption[*]{Illustration of the final state interaction effects in DIS
process (a), initial state interactions in Drell-Yan process (b), and
initial/final state interactions in dijet-correlation process (c). In the
dijet-correlation process, the gauge boson shall couple to any of the
initial/final quarks.}
\label{fig-1}
\end{figure}

We will adopt this model to calculate the DIS, Drell-Yan, and dijet
processes. In particular, we will study the associated quark distribution
functions in the small-$x$ limit and investigate the universality issue. In
Fig.~1, we plot the schematic diagrams for these processes in the scalar-QED
model: (a) for DIS; (b) for Drell-Yan; (c) for dijet-correlation. In the DIS
process (Fig.~1(a)), virtual photon scatters on the scalar quark from the
nucleus target. In our example, the nucleus target has strong coupling with
the Abelian gluon, $g\gg 1$. We need to resum all order gluon exchange
contributions with the nucleus target, for which we have shown in the
diagram. These interactions are referred as final state interactions.
Similarly, we have multi-gluon interaction contribution between the incoming
scalar quark with the nucleus target in the Drell-Yan process as shown in
Fig.~1(b). For the dijet-correlation process in Fig.~1(c), there are both
initial and final state interaction contributions.

The quark distribution functions in the DIS and Drell-Yan processes in this
scalar-QED model have been calculated in the literature. Let us recapture
the main step in these calculations. In the high energy scattering process,
we apply the power counting method to separate short distance physics from
that from long distance. This effectively factorizes the cross section in
terms of parton distributions. In the multi-gluon exchange contributions
illustrated in Fig.~1, the dominant contribution in high energy limit comes
from the parton distribution in nucleus target. For example, in the DIS
process, the leading power contribution to the differential cross section
can be factorized into the quark distribution from nucleus in Fig.~1(a). Of
course, higher order corrections will be taken into account for the gluon
radiation diagrams.

The first step for this factorization is the eikonal approximation, which is
valid in the leading power contribution, i.e., $1/Q^{2}$ for DIS and
Drell-Yan processes. Under this limit, the final state interaction
contribution diagram can be simplified as the eikonal propagator, which can
also be summarized as the gauge link contribution from the associated parton
distribution definition in these two processes. For the dijet-correlation,
since there is no simple definition, we will not seek the gauge link
definition for that, although there has been attempt to do that in the
literature\cite{mulders}. However, we emphasize in the leading power
contribution, we shall be able to obtain the effective parton distributions
in terms of the parton transverse momentum and longitudinal momentum
fraction. We notice that higher order corrections will introduce large
logarithms. To correctly resum these large logarithms, we need to pay
special attention to the transverse momentum dependent parton distributions.
At the current level of this paper, we do not need to worry about this
additional effect.

\begin{figure}[tbp]
\begin{center}
\includegraphics[width=12cm]{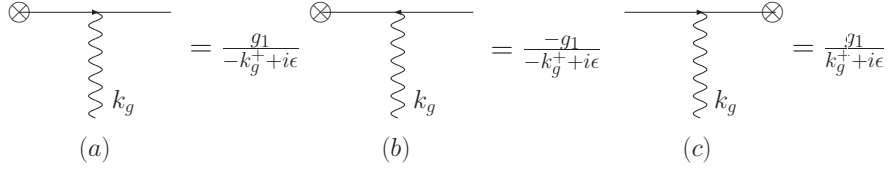}
\end{center}
\caption[*]{Eikonal propagators for the initial/final state interactions in
the scalar-QED model: (a) final state interaction on the scalar quark line;
(b) final state interaction on the scalar antiquark line; (c) initial state
interaction on scalar quark line.}
\label{eik}
\end{figure}

In the scalar-QED model, the eikonal approximation leads to the final state
interaction effect in Fig.~2(a) as,
\begin{equation}
\frac{i}{(k-k_{g})^{2}-m^{2}+i\epsilon }(-ig_{1})(2k-k_{g})\cdot
P_{A}\approx \frac{g_{1}P_{A}^{+}}{-k_{g}^{+}+i\epsilon },  \label{e5}
\end{equation}%
which is the same as that for the fermionic propagator. Similarly, we will
obtain the eikonal propagator for the final state interaction on the scalar
antiquark as illustrated in Fig.~2(b), and the initial state interaction on
the scalar quark in Fig.~2(c). The initial state interaction on the scalar
antiquark will be opposite to that in Fig.~2(c). For the charge $g_{2}$
scalar quarks, we will have the same expressions by replacing $g_{1}$ with $%
g_{2}$ in these diagrams, respectively. In the following calculations, we
will utilize these eikonal approximation for the relevant Feynman diagrams.

\subsection{Universality of Quark Distributions Between SIDIS and Drell-Yan
Processes}

In this subsection, we review the known results\cite{BelJiYua02,
Brodsky:2002ue, Peigne:2002iw} for the quark distributions in the above
described model for the DIS and Drell-Yan lepton pair production processes
in the small-$x$ limit. In particular, we will perform the eikonal
approximation on the final state interactions on the quark line. For the
antiquark line with momentum $p_{2}$, because its phase space is integrated
out, we will keep the full kinematic dependence in these diagrams. The
scalar quark (with charge $g_{1}$) distribution in DIS process can be
written as
\begin{equation}
{\tilde{q}}^{\mathrm{DIS}}(x,q_{\perp })=\frac{x}{32\pi ^{2}}\int \frac{%
dp_{2}^{-}}{p_{2}^{-}}\int \frac{d^{2}k_{\perp }}{(2\pi )^{4}}%
|4P^{+}p_{2}^{-}\sum\limits_{n=1}^{\infty }(gg_{1})^{n}A_{N}^{(n)}(k_{\perp
},k_{\perp }-q_{\perp })|^{2}\ ,  \label{quark}
\end{equation}%
where the first three expansions of the amplitude are found as follows
\begin{eqnarray}
A_{\text{DIS}}^{(1)} &=&\frac{1}{k_{\perp }^{2}+\lambda ^{2}}\left[ \frac{1}{%
D_{1}}-\frac{1}{D_{2}}\right] \ , \\
A_{\text{DIS}}^{(2)} &=&\frac{i}{2!}\int d[1]d[2]\left[ \frac{1}{D_{1}}+%
\frac{1}{D_{2}}-\frac{2}{D_{12}}\right] \ , \\
A_{\text{DIS}}^{(3)} &=&\frac{1}{3!}\int d[1]d[2]d[3]\left[ -\frac{1}{D_{1}}+%
\frac{1}{D_{2}}-\frac{3}{D_{21}}+\frac{3}{D_{12}}\right] \ .
\end{eqnarray}%
For convenience, we have defined the following integral,
\begin{eqnarray}
\int d[1]d[2] &=&\int \frac{d^{2}k_{1\perp }d^{2}k_{2\perp }}{\left( 2\pi
\right) ^{4}}\frac{1}{k_{1\perp }^{2}+\lambda ^{2}}\frac{1}{k_{2\perp
}^{2}+\lambda ^{2}}(2\pi )^{2}\delta ^{(2)}(k_{\perp }-k_{1\perp }-k_{2\perp
})\ ,  \notag \\
\int d[1]d[2]d[3] &=&\int \frac{d^{2}k_{1\perp }d^{2}k_{2\perp
}d^{2}k_{3\perp }}{\left( 2\pi \right) ^{6}}\frac{1}{k_{1\perp }^{2}+\lambda
^{2}}\frac{1}{k_{2\perp }^{2}+\lambda ^{2}}\frac{1}{k_{3\perp }^{2}+\lambda
^{2}}  \notag \\
&&\times (2\pi )^{2}\delta ^{(2)}(k_{\perp }-k_{1\perp }-k_{2\perp
}-k_{3\perp })\ .  \label{3int}
\end{eqnarray}%
We have also defined $D\left( p_{\perp }\right) =2xP^{+}p_{2}^{-}+p_{2\perp
}^{2}+m^{2}$, $D_{1}=D(q_{\perp })$, $D_{2}=D(p_{2\perp })$, $%
D_{1i}=D(q_{\perp }-k_{i\perp })$ and $D_{2i}=D(p_{2\perp }-k_{i\perp })$.
To perform the resummation, we introduce the following Fourier transform,
\begin{equation}
\tilde{A}(R_{\perp },r_{\perp })=\int \frac{d^{2}p_{2\perp }}{(2\pi )^{2}}%
\frac{d^{2}k_{\perp }}{(2\pi )^{2}}A(k_{\perp },p_{2\perp })e^{-i\vec{%
k_{\perp }}\cdot R_{\perp }-i\vec{p}_{2\perp }\cdot \vec{r}_{\perp }}\ .
\end{equation}%
By applying this Fourier transform, we obtain,
\begin{eqnarray}
\tilde{A}_{\text{DIS}}^{(1)} &=&-V(r_{\perp })W(r_{\perp },R_{\perp })\ , \\
\tilde{A}_{\text{DIS}}^{(2)} &=&+\frac{i}{2!}V(r_{\perp })W^{2}(r_{\perp
},R_{\perp })\ , \\
\tilde{A}_{\text{DIS}}^{(3)} &=&+\frac{1}{3!}V(r_{\perp })W^{3}(r_{\perp
},R_{\perp })\ ,
\end{eqnarray}%
where\ the functions $V$ and $W$ are defined as
\begin{eqnarray}
V(r_{\perp }) &=&\frac{1}{2\pi }K_{0}(M_{0}r_{\perp })\ , \\
W(r_{\perp },R_{\perp }) &=&\frac{1}{2\pi }\log \left( \frac{|\vec{R}_{\perp
}+\vec{r}_{\perp }|}{R_{\perp }}\right) \ ,
\end{eqnarray}%
with $M_{0}^{2}=2xP^{+}p_{2}^{-}+m^{2}$, and $K_{0}$ is the Bessel function.
Clearly, we can see that the above expansion comes from the following
exponential form,
\begin{equation}
\tilde{A}_{\text{DIS}}=iV(r_{\perp })\left[ 1-e^{-igg_{1}W(r_{\perp
},R_{\perp })}\right] \ .
\end{equation}%
Substituting the above result into the quark distribution expression, we
will obtain the TMD quark distribution in the DIS process,
\begin{eqnarray}
\tilde{q}^{\mathrm{DIS}}(x,q_{\perp }) &=&\frac{xP^{+2}}{8\pi ^{4}}\int
dp_{2}^{-}p_{2}^{-}\int d^{2}R_{\perp }d^{2}r_{\perp }d^{2}r_{\perp
}^{\prime }e^{-i\vec{q}_{\perp }\cdot (\vec{r}_{\perp }-\vec{r}_{\perp
}^{\prime })}V(r_{\perp })V(r_{\perp }^{\prime })  \notag \\
&&\times \left[ 1-e^{-igg_{1}W(r_{\perp }^{{}},R_{\perp }^{{}})}\right] %
\left[ 1-e^{igg_{1}W(r_{\perp }^{\prime },R_{\perp })}\right] .
\end{eqnarray}%
In Sec.III, we will discuss how to translate this result into the fermion
quark distribution function at small-$x$ for this process in QCD and compare
to the color-dipole/color glass condensate formalism.

Similarly, one can calculate the quark distribution\cite{Peigne:2002iw} in
the Drell-Yan lepton pair production process. The definition of the quark
distribution has the same expression as in Eq.~(\ref{quark}). The amplitudes
of first three orders are found to be
\begin{eqnarray}
A_{\text{DY}}^{(1)} &=&\frac{1}{k_{\perp }^{2}+\lambda ^{2}}\left[ \frac{1}{%
D_{1}}-\frac{1}{D_{2}}\right] \text{ }, \\
A_{\text{DY}}^{(2)} &=&\frac{i}{2!}\left[ \frac{1}{D_{1}}-\frac{1}{D_{2}}%
\right] \int d[1]d[2]\ , \\
A_{\text{DY}}^{(3)} &=&\frac{-1}{3!}\left[ \frac{1}{D_{1}}-\frac{1}{D_{2}}%
\right] \int d[1]d[2]d[3]\ .
\end{eqnarray}%
It is easier to do the resummation when we perform the Fourier transform,
for which we have
\begin{eqnarray}
\tilde{A}_{\text{DY}}^{(1)} &=&-V(r_{\perp })\left[ G(R_{\perp })-G(\vec{R}%
_{\perp }+\vec{r}_{\perp })\right] \ , \\
\tilde{A}_{\text{DY}}^{(2)} &=&-\frac{i}{2!}V(r_{\perp })\left[
G^{2}(R_{\perp })-G^{2}(\vec{R}_{\perp }+\vec{r}_{\perp })\right] \ , \\
\tilde{A}_{\text{DY}}^{(3)} &=&+\frac{1}{3!}V(r_{\perp })\left[
G^{3}(R_{\perp })-G^{3}(\vec{R}_{\perp }+\vec{r}_{\perp })\right] \ ,
\end{eqnarray}%
where $G(R_{\perp })=\frac{1}{2\pi }K_{0}(\lambda R_{\perp })$. The all
order resummation leads to
\begin{equation}
\tilde{A}_{\text{DY}}=iV(r_{\perp })e^{igg_{1}G(R_{\perp })}\left[
1-e^{igg_{1}\left( G(\vec{R}_{\perp }+\vec{r}_{\perp })-G(R_{\perp })\right)
}\right] \ .
\end{equation}%
There is infrared divergence when the gluon mass goes to zero $\lambda
\rightarrow 0$. However, this infrared divergence cancels out when we
calculate the quark distribution. Furthermore, taking the limit $\lambda
\rightarrow 0$, we also have $-G(\vec{R}_{\perp }+\vec{r}_{\perp
})+G(R_{\perp })\rightarrow W(r_{\perp },R_{\perp })$, and we can find that
the quark distribution is universal between DIS and Drell-Yan processes
\begin{equation}
\tilde{q}^{\mathrm{DY}}(x,q_{\perp })=\tilde{q}^{\mathrm{DIS}}(x,q_{\perp
})\ .
\end{equation}%
This universality is also guaranteed by the time-reversal invariance. The
quark distributions in these two processes can be connected through
time-reversal transformation, and time-reversal invariance can show that
they are the same. However, this will not be the case when we compare the
photon-jet correlation in Drell-Yan type process and dijet-correlation in
DIS type process. We will carry out these calculations in future studies.

\subsection{Non-universality of Quark Distribution Between Dijet-Correlation
and SIDIS/DY Processes}

To study the TMD scalar quark distribution in hadronic processes (e.g., $pA$
collisions), we introduce the dijet-correlation process as illustrated in
Fig.~1(c),
\begin{equation}
p+A\rightarrow \mathrm{Jet1}+\mathrm{Jet2}+X\ ,
\end{equation}%
where the transverse momenta of these two jets are similar in size but
opposite to each other in direction. For simplicity, we focus on the
partonic channel $qq^{\prime }\rightarrow qq^{\prime }$ for now in this
paper and study the quark distribution. We expect that similar conclusions
can be drawn as to other partonic channels and thus to the gluon
distributions as well. In the ideal case, these two jets are produced
back-to-back. However, the gluon radiation and intrinsic transverse momenta
of the initial partons induce an imbalance between them. We are particularly
interested in the kinematic region that the imbalance $\vec{q}_{\perp }=\vec{%
P}_{1\perp }+\vec{P}_{2\perp }$ is much smaller than the transverse momentum
of the individual jet, namely, $|\vec{q}_{\perp }|\ll |\vec{P}_{1\perp
}|\sim |\vec{P}_{2\perp }|$, which also corresponds to the kinematics in the
STAR measurements. Only in this region, can the intrinsic transverse
momentum have significant effects. Since there are two incoming partons,
both intrinsic transverse momenta can affect the imbalance between the two
jets. For large nucleus and small-$x$, the dominant contribution should come
from the intrinsic transverse momentum of the parton from the nucleus, which
we label as $q_{\perp }$ in Fig.~1(a). In the following, we will focus on
this contribution.

Since we are interested in studying the final state interaction effects on
the parton distribution of the nucleus, for convenience, we choose the
projectile as a single scalar quark with charge $g_{2}$, which differs from
the charge of the scalar quark from the target nucleus, $g_{1}$. In
addition, we assume that the Abelian gluon attaches to the target nucleus
with an effective coupling $g$ which is much larger than $g_{2}$ or $g_{1}$.
All the partons in this calculation are set to be scalars with a mass $m$.
The coupling $g_{2}$ being different from $g_{1}$ is to show the dependence
of the parton distribution on the initial/final state interactions
associated with the incoming parton. If the dependence on $g_{2}$ remains
for the nucleus parton distributions, they are not universal~\cite%
{Collins:2007nk,{Vogelsang:2007jk}}.

\begin{figure}[tbp]
\begin{center}
\includegraphics[width=4cm]{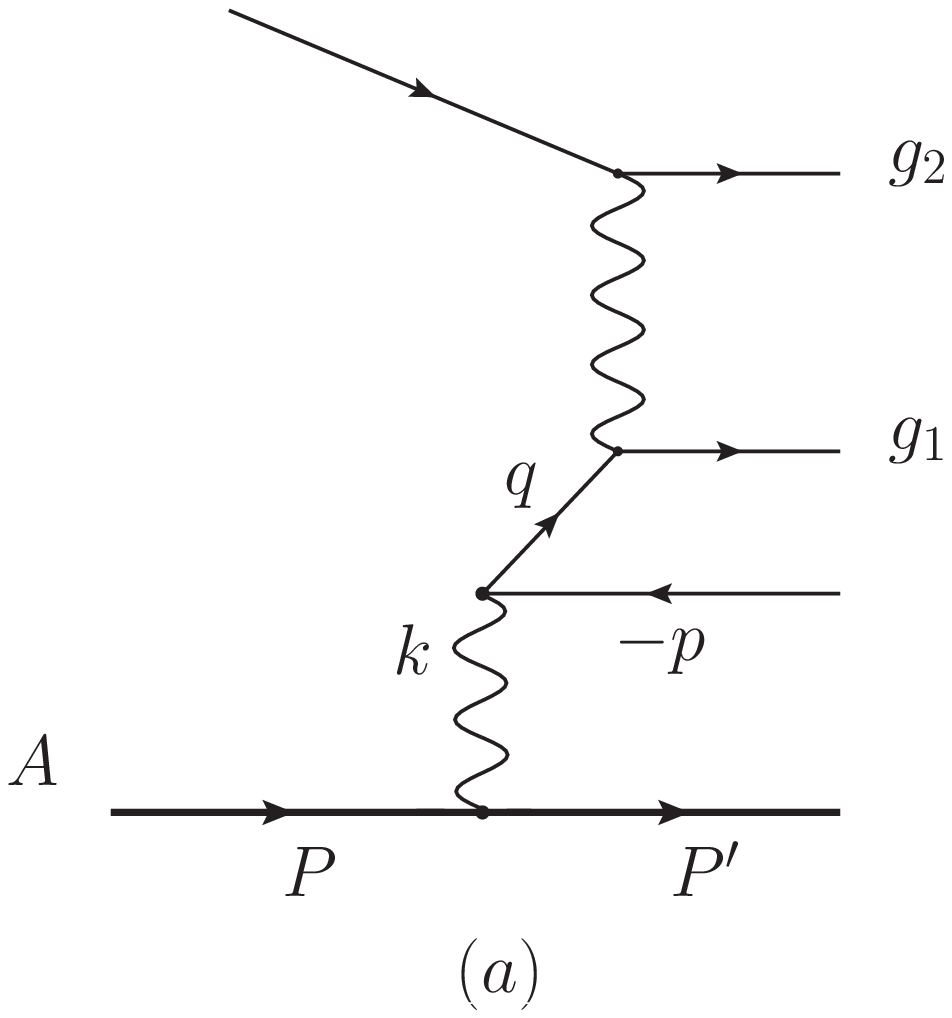}\hfill \includegraphics[width=4cm]{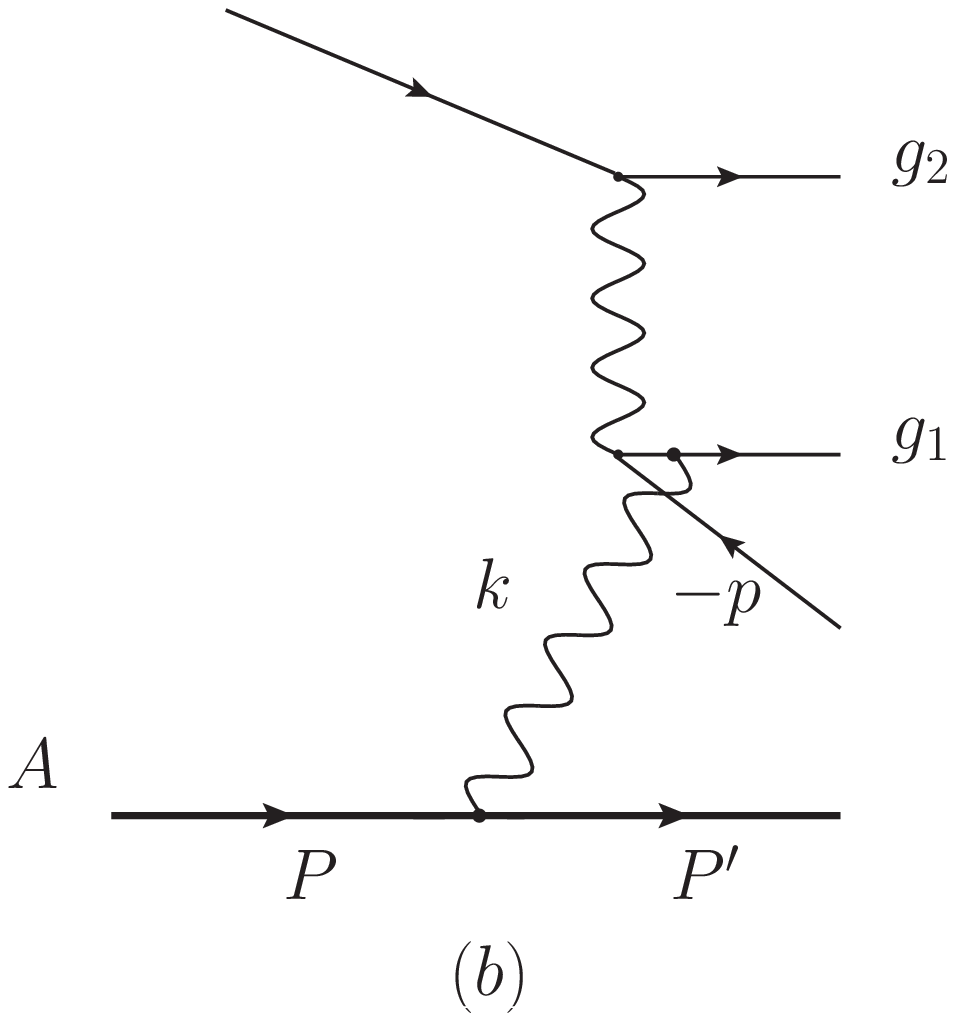}%
\hfill 
\includegraphics[width=4cm]{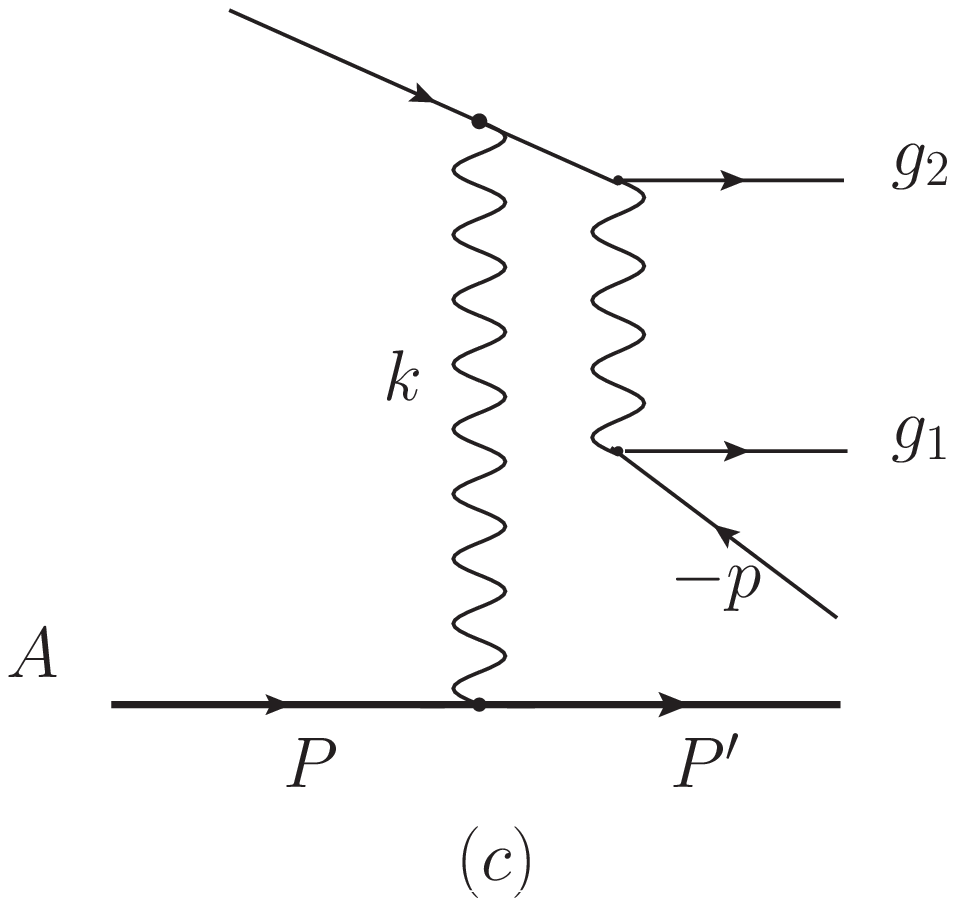}\hfill \includegraphics[width=4cm]{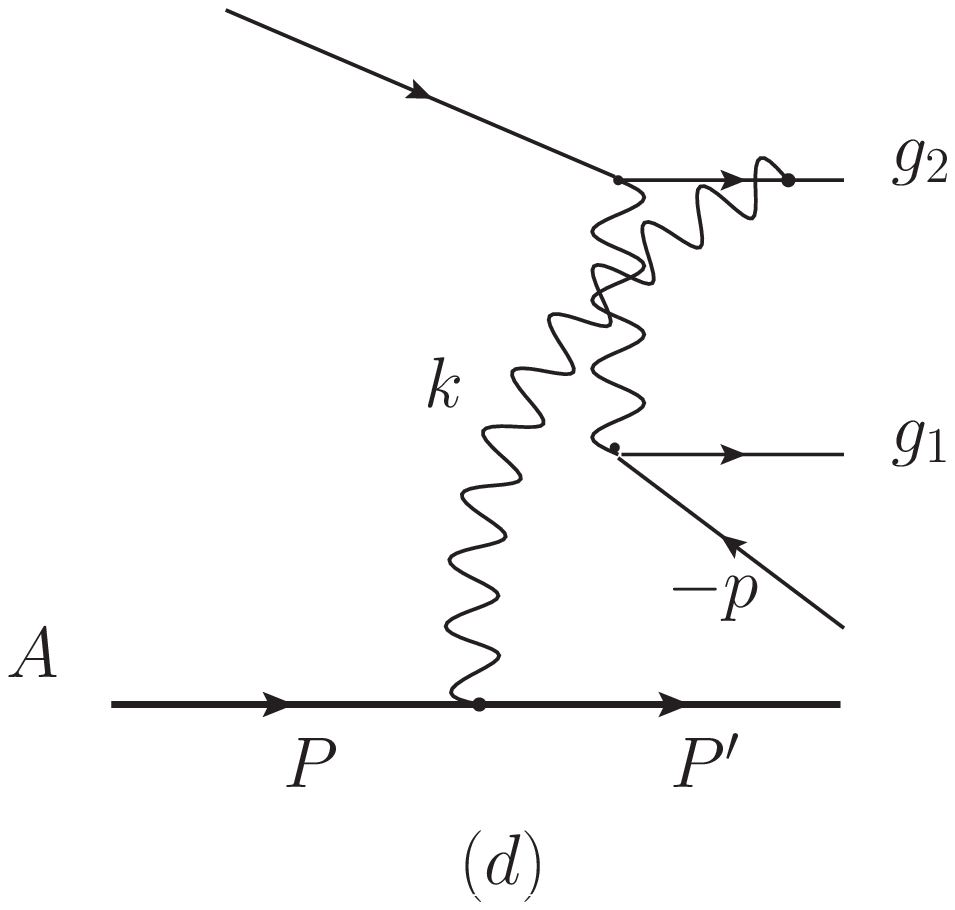}%
\hfill
\end{center}
\caption[*]{Lowest-order graphs for di-jet production in a hadron-hadron
collision in the small-x limit. In these graphs, there is one soft gluon
exchange with momentum $k$ in addition to the hard gluon exchange.}
\label{fig1}
\end{figure}

In Fig.~\ref{fig1}, we plot the lowest-order graphs containing one soft
gluon exchange with the momentum $k$. Following the discussions above, we
keep the low transverse momentum approximation in terms of $q_{\perp
}/P_{1\perp }$ ($q_{\perp }/P_{2\perp }$) by applying the power counting
method~\cite{qvy-short}. Again, the important simplification is the eikonal
approximation, which replaces the gluon attachment to the initial and final
state partons with the eikonal propagator and vertex. After taking the
leading order contributions, we find that the $q_{\perp }$ dependence of
these diagrams can be cast into an effective quark distribution~\cite%
{qvy-short}, which takes the following form,
\begin{equation}
\tilde{q}\left( x,q_{\perp }\right) =\frac{x}{32\pi ^{2}}\int \frac{dp^{-}}{%
p^{-}}\frac{d^{2}k_{\perp }}{\left( 2\pi \right) ^{4}}(4P^{+}p^{-})^{2}\left%
\vert A^{(tot)}\left( k,p\right) \right\vert ^{2}\ ,  \label{dijetqd}
\end{equation}%
with $p_{\perp }=k_{\perp }-q_{\perp }$. Here, the hard partonic part
depending on the hard momentum scale $P_{i\perp }$ has been separated from
the above quark distribution in the differential cross section~\cite%
{qvy-short}. This separation is only possible at the leading power
contribution of $q_{\perp }/P_{i\perp }$. The contributions from Fig.~2 can
be written as,
\begin{equation}
A^{(1)}\left( k,p\right) =gg_{1}\frac{1}{k_{\perp }^{2}+\lambda ^{2}}\left[
\frac{1}{D_{1}}-\frac{1}{D_{2}}\right] +gg_{2}\frac{1}{k_{\perp
}^{2}+\lambda ^{2}}\frac{1}{2p^{-}}\left[ \frac{1}{-k^{+}+i\epsilon }+\frac{1%
}{k^{+}+i\epsilon }\right] \ ,  \label{oneglu}
\end{equation}%
where $D_{i}$ follow the definitions introduced above. The first and second
terms in the first square bracket correspond to Fig.~\ref{fig1} (a) and Fig.~%
\ref{fig1} (b), respectively. Fig.~\ref{fig1} (c) and Fig.~\ref{fig1} (d)
yield the contributions as shown in the last term in Eq.~(\ref{oneglu}). It
is not hard to see that these two contributions simply cancels since the sum
is proportional to $2\pi i\delta (k^{+})$ while $k^{+}\neq 0$. This means
that at the leading order in the coupling constant the dependence on $g_{2}$
drops out, which will however change at higher orders. Thus, one gets $%
A^{(1)}\left( k,p\right) =gg_{1}\frac{1}{k_{\perp }^{2}+\lambda ^{2}}\left[
\frac{1}{D_{1}}-\frac{1}{D_{2}}\right] $.

\begin{figure}[tbp]
\begin{center}
\includegraphics[width=5cm]{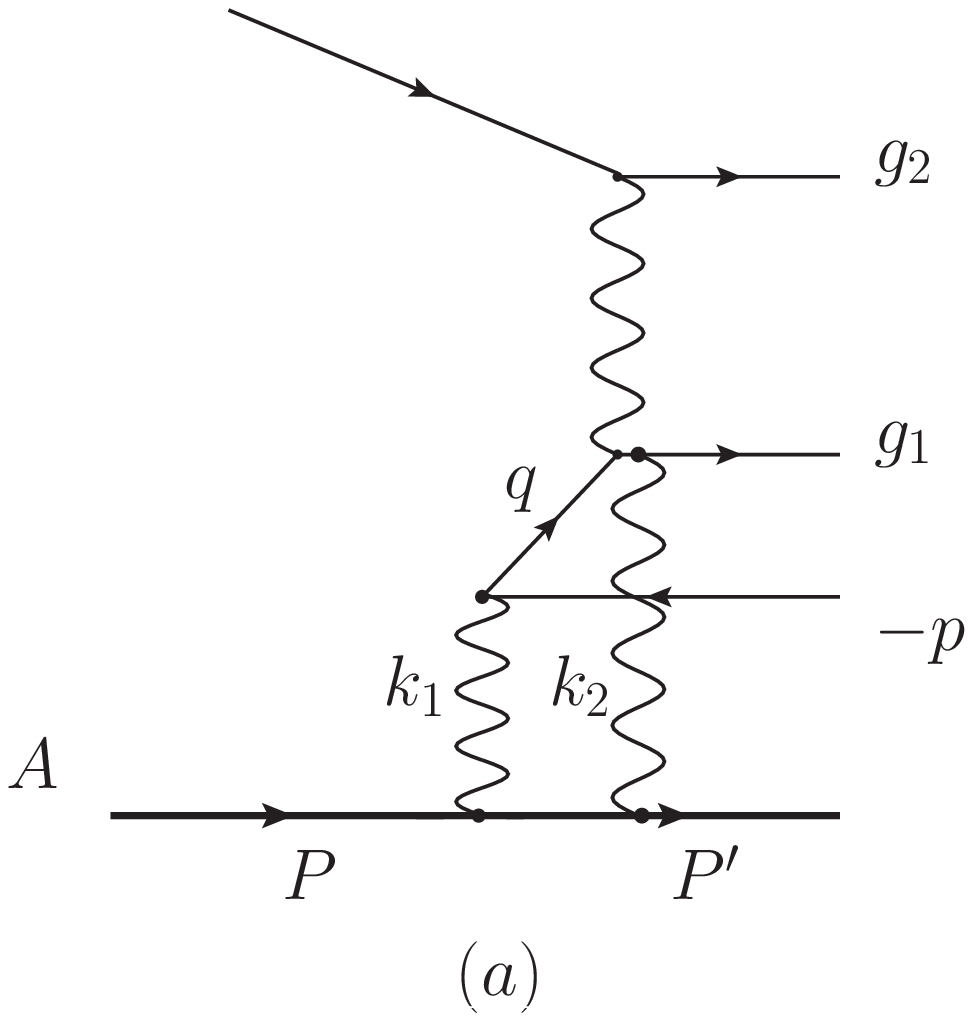}
\includegraphics[width=5cm]{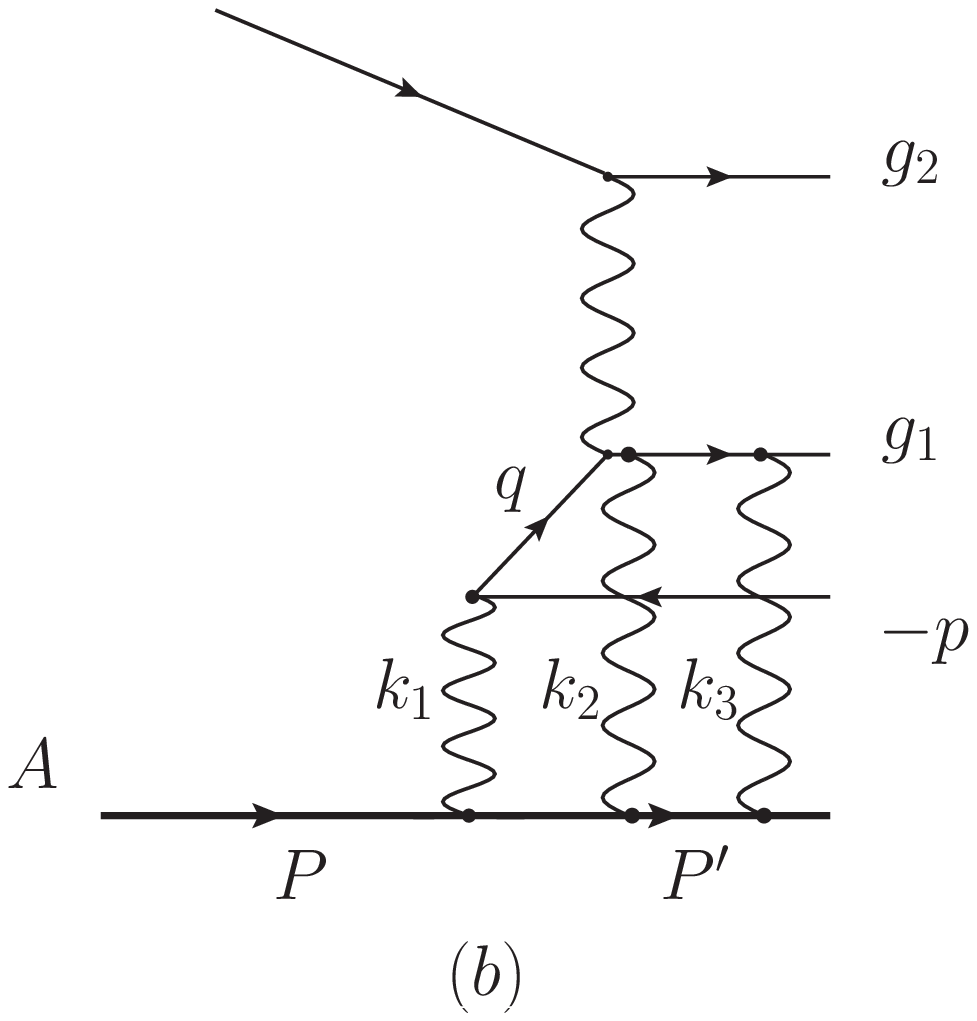}
\end{center}
\caption[*]{Example diagrams for two (a) and three (b) gluons exchanges,
where the gluons can attach all charged particles in the upper part of the
diagrams to the nucleus target.}
\label{fig2}
\end{figure}

\begin{figure}[tbp]
\begin{center}
\includegraphics[width=12cm]{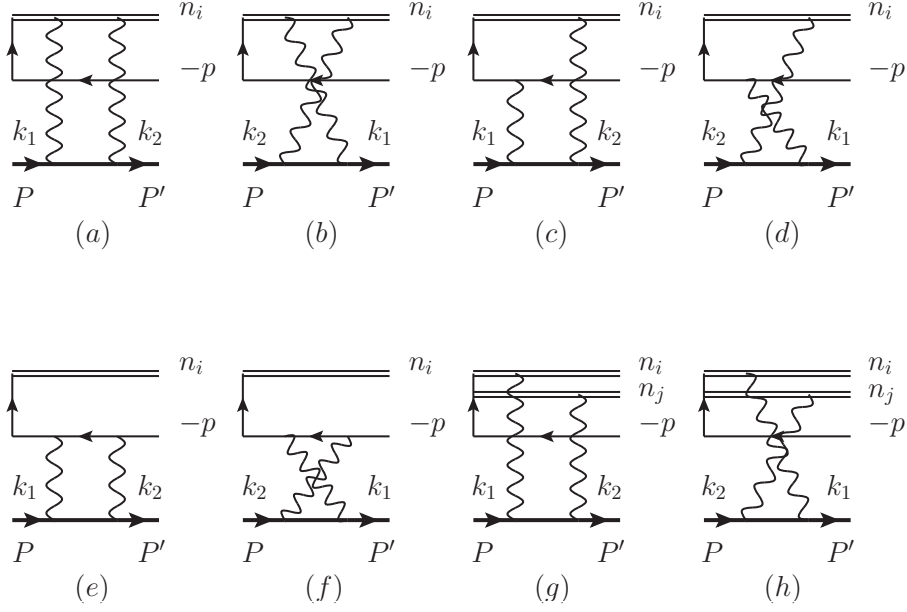}
\end{center}
\caption[*]{Two-gluon-exchange contributions to the quark distribution. $%
i=1,2,3$ for each $n_i$ in each graph with $n_1=\frac{g_1}{-k^{+}+i \protect%
\epsilon}$, $n_2=\frac{g_2}{-k^{+}+i\protect\epsilon}$ and $n_3=\frac{g_2}{
k^{+}+i\protect\epsilon}$. $i\neq j$ for $n_{i,j}$ is implied in graphs (g)
and (h).}
\label{fig2-1}
\end{figure}

At the next-to-leading order and the $g^{3}$ order, there are 20 and 120
graphs in total in covariant gauge, respectively. We show one of these
graphs as an example in Fig.~(\ref{fig2}) (a) and (b), respectively.
Additional diagrams can be obtained by attaching the gluons to all incoming
and outgoing scalar quarks. We organize our calculations according to these
attachments. Again, the eikonal approximation discussed at the beginning of
this sections will be utilized in the evaluations of these diagrams. For
example, by applying this approximation, we replace the final state
QED-gluon interactions with the scalar quark line with the eikonal
propagators and vertices associated with $g_1$ coupling, which we label with
$n_1$ in Fig.~\ref{fig2-1}. Similar approximations are made for all other
diagrams, with $n_2$ and $n_3$ representing final and initial state
QED-gluon interaction with the scalar quark line with charge $g_2$. We
summarize those diagrams in Fig.~(\ref{fig2-1}), where every graph except
for (e) and (f) represents 3 graphs and total of 20 graphs contribute.

Among those 20 graphs, only a few sets of graphs give non-vanishing
contributions. To better explain the calculation, let us first evaluate
Fig.~(\ref{fig2-1}) (a) and (b) as an example. Fig.~(\ref{fig2-1})(a) and
Fig.~(\ref{fig2-1})(b) with $n_{i}=n_{1}$ give the following contribution%
\begin{eqnarray}
A_{a+b}^{\left( 2,i=1\right) } &=&\frac{ig^{2}g_{1}^{2}\left( 2P^{+}\right)
^{2}}{4P^{+}p^{-}}\int \frac{\text{d}^{4}k_{2}}{\left( 2\pi \right) ^{4}}%
\frac{1}{k_{1}^{2}-\lambda ^{2}}\frac{1}{k_{2}^{2}-\lambda ^{2}}\frac{1}{%
k_{2}^{+}-i\epsilon }\frac{1}{k^{+}-i\epsilon }  \notag \\
&&\times \left[ \frac{1}{\left( P-k_{1}\right) ^{2}-m^{2}+i\epsilon }+\frac{1%
}{\left( P-k_{2}\right) ^{2}-m^{2}+i\epsilon }\right] .
\end{eqnarray}%
After integrating over d$k_{2}^{-}$ and d$k_{2}^{+}$, one finds that the
subtotal contribution reads
\begin{equation}
A_{a+b}^{\left( 2,i\right) }=\frac{ig^{2}g_{1}^{2}}{2}\int \frac{\text{d}%
^{2}k_{2\perp }}{\left( 2\pi \right) ^{2}}\frac{1}{k_{1\perp }^{2}+\lambda
^{2}}\frac{1}{k_{2\perp }^{2}+\lambda ^{2}}\frac{1}{D\left( p_{\perp
}\right) }.
\end{equation}%
It is straightforward to find that the contributions from $n_{2}$ and $n_{3}$
cancels and $g_{2}^{2}$ contribution vanishes due to the same reason we
explained under Eq.~(\ref{oneglu}).

As to the graphs illustrated in Fig.~(\ref{fig2-1})(c) and Fig.~(\ref{fig2-1}%
)(d), we find that the contribution from $n_{3}$ vanishes due to vanishing
contour integral of d$k_{2}^{+}$. Thus, the subtotal contribution of Fig.~(%
\ref{fig2-1})(c) and Fig.~(\ref{fig2-1})(d) is
\begin{eqnarray}
A_{c+d}^{\left( 2,i\right) } &=&\frac{ig^{2}\left(
g_{1}^{2}+g_{1}g_{2}\right) \left( 2P^{+}\right) ^{2}\left( -2p^{-}\right) }{%
4P^{+}p^{-}}\int \frac{\text{d}^{4}k_{2}}{\left( 2\pi \right) ^{4}}\frac{1}{%
k_{1}^{2}-\lambda ^{2}}\frac{1}{k_{2}^{2}-\lambda ^{2}}\frac{1}{%
k_{2}^{+}-i\epsilon }  \notag \\
&&\times \frac{1}{\left( p-k_{1}\right) ^{2}-m^{2}+i\epsilon }\left[ \frac{1%
}{\left( P-k_{1}\right) ^{2}-m^{2}+i\epsilon }+\frac{1}{\left(
P-k_{2}\right) ^{2}-m^{2}+i\epsilon }\right] ,
\end{eqnarray}%
with $k=k_{1}+k_{2}$. After integrating over d$k_{2}^{-}$ and d$k_{2}^{+}$,
one finds
\begin{equation}
A_{c+d}^{\left( 2,i\right) }=-\frac{ig^{2}\left( g_{1}^{2}+g_{1}g_{2}\right)
}{2}\int \frac{\text{d}^{2}k_{2\perp }}{\left( 2\pi \right) ^{2}}\frac{1}{%
k_{1\perp }^{2}+\lambda ^{2}}\frac{1}{k_{2\perp }^{2}+\lambda ^{2}}\left[
\frac{1}{D(p_{\perp }-k_{1\perp })}+\frac{1}{D(p_{\perp }-k_{2\perp })}%
\right] .
\end{equation}%
Similarly, we obtain
\begin{equation}
A_{e+f}^{\left( 2,i\right) }=-\frac{ig^{2}g_{1}^{2}}{2}\int \frac{\text{d}%
^{2}k_{2\perp }}{\left( 2\pi \right) ^{2}}\frac{1}{k_{1\perp }^{2}+\lambda
^{2}}\frac{1}{k_{2\perp }^{2}+\lambda ^{2}}\frac{1}{D(q_{\perp })},
\end{equation}%
and
\begin{equation}
A_{g+h}^{\left( 2,i\right) }=ig^{2}g_{1}g_{2}\int \frac{\text{d}%
^{2}k_{2\perp }}{\left( 2\pi \right) ^{2}}\frac{1}{k_{1\perp }^{2}+\lambda
^{2}}\frac{1}{k_{2\perp }^{2}+\lambda ^{2}}\frac{1}{D\left( p_{\perp
}\right) }.
\end{equation}

The total contributions are
\begin{equation}
A^{(2)}\left( k,p\right) =\frac{i}{2}g^{2}\int d[1]d[2]\left\{ g_{1}^{2}%
\left[ \frac{1}{D_{1}}+\frac{1}{D_{2}}-\frac{1}{D_{21}}-\frac{1}{D_{22}}%
\right] +g_{1}g_{2}\left[ \frac{2}{D_{2}}-\frac{2}{D_{21}}\right] \right\} \
,
\end{equation}%
where $\int d[1]d[2]$ stands for the same definition as in Eq.~(\ref{3int}),
$D_{1i}=D(q_{\perp }-k_{i\perp })$ and $D_{2i}=D(p_{\perp }-k_{i\perp })$.
Clearly, the second order result shows a dependence on $g_{2}$. However, in
the amplitude squared calculation for the quark distribution Eq.~(\ref%
{dijetqd}), the $g_{2}$ dependence from $A^{(1)}{A^{(2)}}^{\ast }$ is
canceled out by its complex conjugate because $A^{(1)}$ is real while ${%
A^{(2)}}$ is purely imaginary. The leading order contribution to $g_{2}$
comes from $\left\vert A^{(2)}\left( k,p\right) \right\vert ^{2}$ and $%
A^{(1)}{A^{(3)}}^{\ast }+A^{(3)}{A^{(1)}}^{\ast }$. Therefore, to see the
residue dependence on $g_{2}$, we need to carry out the calculation of the
amplitude up to order $g^{3}$.

At the $g^{3}$ order, there are 120 diagrams in total with three soft
gluon-exchange (see e.g., Fig.~\ref{fig2} (b)), including all possible
permutations of the attachments of these three gluons to the target nucleus.
Let us take Fig.~\ref{fig2} (b) as an example together with the other 5
crossing diagrams. The corresponding contribution is%
\begin{equation}
I_{1}^{\left( 3\right) }=-\frac{g^{3}g_{1}^{3}}{2}\int \frac{\text{d}%
^{2}k_{1\perp }\text{d}^{2}k_{2\perp }}{\left( 2\pi \right) ^{4}}\frac{1}{%
k_{1\perp }^{2}+\lambda ^{2}}\frac{1}{k_{2\perp }^{2}+\lambda ^{2}}\frac{1}{%
k_{3\perp }^{2}+\lambda ^{2}}\frac{1}{D\left( p_{\perp }-k_{1\perp }\right) }%
.
\end{equation}%
In reaching above result, we have used the following two integrals:%
\begin{eqnarray}
&&\int \frac{\text{d}k_{1}^{-}\text{d}k_{2}^{-}\text{d}k_{3}^{-}}{\left(
2\pi \right) ^{2}}\delta (k^{-}-k_{1}^{-}-k_{2}^{-}-k_{3}^{-})  \notag \\
&&\times \left[ \frac{i}{\left( P-k_{1}\right) ^{2}-m^{2}+i\epsilon }\frac{i%
}{\left( P-k_{1}-k_{2}\right) ^{2}-m^{2}+i\epsilon }+\text{Crossing Diagrams}%
\right]  \notag \\
&=&\frac{-1}{4P^{+2}}\int \frac{\text{d}k_{2}^{-}\text{d}k_{3}^{-}}{\left(
2\pi \right) ^{2}}\frac{k^{-}}{\left( k^{-}-k_{2}^{-}-k_{3}^{-}-i\epsilon
\right) \left( k_{2}^{-}-i\epsilon \right) \left( k_{3}^{-}-i\epsilon
\right) }=\frac{1}{4P^{+2}}.
\end{eqnarray}%
and
\begin{eqnarray}
&&\int \frac{\text{d}k_{1}^{+}\text{d}k_{2}^{+}\text{d}k_{3}^{+}}{\left(
2\pi \right) ^{2}}\delta (k^{+}-k_{1}^{+}-k_{2}^{+}-k_{3}^{+})  \notag \\
&&\times \left[ \frac{i}{\left( p-k_{1}\right) ^{2}-m^{2}+i\epsilon }\frac{i%
}{-k_{3}^{+}+i\epsilon }\frac{i}{-k_{2}^{+}-k_{3}^{+}+i\epsilon }\right]
\notag \\
&=&-\frac{i}{2}\frac{1}{D\left( p_{\perp }-k_{1\perp }\right) }.
\end{eqnarray}

Summing up all these graphs, we obtain the three gluon exchange amplitude,
\begin{eqnarray}
A^{(3)}\left( k,p\right) &=&\frac{1}{3!}g^{3}\int d[1]d[2]d[3]\left\{
g_{1}^{3}\left[ \frac{1}{D_{2}}-\frac{1}{D_{1}}+\frac{3}{D_{13}}-\frac{3}{%
D_{21}}\right] \right.  \notag \\
&&\left. +g_{1}^{2}g_{2}\left[ \frac{3}{D_{2}}+\frac{3}{D_{13}}-\frac{3}{%
D_{21}}-\frac{3}{D_{22}}\right] +g_{1}g_{2}^{2}\left[ \frac{3}{D_{2}}-\frac{3%
}{D_{21}}\right] \right\} \ ,
\end{eqnarray}%
where $\int d[1]d[2]d[3]$ follows the same definition as in Eq.~(\ref{3int}%
). Again, we see the dependence on $g_{2}$ in the second and third terms. An
important cross check of these results is that, if we set $g_{2}=-g_{1}$,
there is effectively no charge flow in the final state, and the quark
distribution is identical to that in the Drell-Yan process in the same
model. Applying $g_{2}=-g_{1}$, we can easily see that indeed we reproduce
those calculated in Ref.~\cite{Peigne:2002iw}. Also, by setting $g_{2}=0$,
we can recover the DIS amplitudes.

With the amplitude calculated up to $\mathcal{O}\left( g^{3}\right) $, we
are able to check the dependence on $g_{2}$ for the parton distribution in
Eq.~(\ref{dijetqd}). Substituting the above amplitudes into Eq.~(\ref%
{dijetqd}), we find that the $g_{2}$ dependence still remains up to order $%
g^{4}$. If we drop all $g_{2}$ terms in these results, we obtain the quark
distribution in DIS in the same model~\cite{Brodsky:2002ue,BelJiYua02}. This
clearly shows that the TMD quark distribution $\tilde{q}(x,q_{\perp })$ is
not universal.

This non-universality is better illustrated when we sum up all order
multi-gluon exchange contributions. To do that, we introduce the following
Fourier transform~\cite{Brodsky:2002ue},
\begin{equation}
A\left( R,r\right) =\int \frac{d^{2}k_{\perp }}{\left( 2\pi \right) ^{2}}%
\frac{d^{2}p_{\perp }}{\left( 2\pi \right) ^{2}}e^{-ik_{\perp }\cdot
R_{\perp }-ip_{\perp }\cdot r_{\perp }}A\left( k,p\right) \ .
\end{equation}%
From the Fourier transforms of $A^{(1,2,3)}(k,p)$, we can easily see that
they follow the expansion of an exponential form,
\begin{equation}
A^{(tot)}\left( R,r\right) =\sum_{n=1}^{\infty }A^{\left( n\right) }\left(
R,r\right) =iV\left( r_{\perp }\right) \left\{ 1-e^{igg_{1}\left[ G\left(
R_{\perp }+r_{\perp }\right) -G\left( R_{\perp }\right) \right] }\right\}
e^{-igg_{2}G\left( R_{\perp }\right) }\ ,
\end{equation}%
where $G(R_{\perp })=K_{0}\left( \lambda R_{\perp }\right) /2\pi $ and $%
V(r_{\perp })=K_{0}\left( Mr_{\perp }\right) /2\pi $ with $%
M^{2}=2xP^{+}p^{-}+m^{2}$. In the above result, the $g_{2}$-dependence seems
to only appear as a phase which may not lead to a physics consequence.
However, because the transverse momentum $q_{\perp }$ is conjugate to the
coordinate variable difference $R_{\perp }$-$r_{\perp }$, this phase will
lead to a non-universal contribution to the quark distribution as defined in
Eq.~(\ref{dijetqd}). Here we can identify the factor $e^{-igg_{i}G\left(
R_{\perp }\right) }$ as a Wilson line which essentially resums soft
interactions between quarks and the target nucleus.

Therefore, the all order result reads,
\begin{eqnarray}
\tilde{q}\left( x,q_{\perp }\right)  &\!\!=\!\!&\frac{xP^{+2}}{8\pi ^{4}}%
\int dp^{-}p^{-}\int d^{2}R_{\perp }d^{2}R_{\perp }^{\prime }d^{2}r_{\perp
}d^{2}r_{\perp }^{\prime }\delta ^{\left( 2\right) }\left( R_{\perp
}+r_{\perp }-R_{\perp }^{\prime }-r_{\perp }^{\prime }\right)   \notag \\
&&\times e^{-iq_{\perp }\cdot \left( r_{\perp }-r_{\perp }^{\prime }\right)
}e^{-igg_{2}\left( G(R_{\perp })-G(R_{\perp }^{\prime })\right) }V\left(
r_{\perp }\right) V\left( r_{\perp }^{\prime }\right)  \nonumber \\
&&\times \left\{ 1-e^{igg_{1}\left[ G\left( R_{\perp }^{{}}+r_{\perp
}^{{}}\right) -G\left( R_{\perp }\right) \right] }\right\} \left\{
1-e^{-igg_{1}\left[ G\left( R_{\perp }^{\prime }+r_{\perp }^{\prime }\right)
-G\left( R_{\perp }^{\prime }\right) \right] }\right\} \ ,  \label{diq}
\end{eqnarray}%
This TMD quark distribution is clearly different from that calculated in DIS
in the same model~\cite{Brodsky:2002ue,{BelJiYua02}}. In other words, TMD
quark distributions are not universal. It is interesting to notice that the $%
g_{2}$ dependence disappears after the integration over the transverse
momentum. This is consistent with the universality for the integrated parton
distributions~\cite{BelJiYua02,{Collins:2007nk},Vogelsang:2007jk}. In spite
of the non-universality, we expect that one can still reach an effective TMD
factorization formula for $pA$ collisions by absorbing all the violation
effects into the parton distributions as in Eq.~(\ref{diq}).

It has been argued that the light-cone gauge may simplify the factorization
property for the hard scattering processes. For example, if we choose the
advanced boundary condition for the gauge potential in light-cone gauge, the
wave function of hadrons contain the final state interaction effects~\cite%
{BelJiYua02,{Brodsky:2010vs}}. However, as we showed in the above
calculations, this does not help to resolve the $g_{2}$-dependence in the
quark distribution in the dijet correlation in hadronic process due to the
presence of both initial and final state interactions. In other words, the
quark distribution from the nucleus target has to contain the interaction
with the incoming (outgoing) quark with charge $g_{2}$, which can not be
solely included into the wave function of the nucleus target.

\section{Extension to the Quark Distributions at Small-$x$ in CGC}

In this section, we extend the previous calculation to the quark
distribution functions for a large nucleus at small-$x$ in QCD, by
calculating the similar resummation effects due to initial and final state
interactions. As an example, we follow the McLerran-Venugopalan (MV) Model~\cite%
{McLerran:1993ni}. The MV model describes high density gluon distribution in a
relativistic large nucleus by solving the classical Yang-Mills equation. An
effective theory, called the Color Glass Condensate (CGC), is developed to
study the high density physics in QCD in a systematic manner~\cite%
{Iancu:2003xm}. It is equivalent to the saturation picture which is based on
the color dipole model\cite{Mueller:1993rr,Mueller:1999yb} in terms of
quantitatively describing the parton saturation at small-$x$. In our
following discussion, we will focus on the CGC formalism since it is close
to what we have used above in the QED model.

Following the CGC formalism, the target nucleus is treated as a collection
of randomly distributed color sources $\rho _{a}\left( z^{-},z_{\perp
}\right) $ which generate soft classical gluon fields. Similar to what
happens in above QED model, where we obtain the Wilson line $e^{-igg_{i}%
\left[ G\left( x_{\perp }\right) \right] }$, the soft interactions (i.e.,
the soft gluon exchanges) between the nucleus and a relativistic quark can
also be resummed into a Wilson line which reads%
\begin{equation}
U\left( x_{\perp }\right) =\text{\textit{T}}\mathit{\exp }\left[
-igg_{1}\int {d}z^{-}{d}^{2}z_{\perp }G\left( x_{\perp }-z_{\perp }\right)
\rho _{a}\left( z^{-},z_{\perp }\right) t^{a}\right] ,
\end{equation}%
where $t^{a}$ is the $SU(3)$ color matrix in the fundamental representation
and the two-dimensional propagator $G\left( x_{\perp }-z_{\perp }\right) $
is the same as the one we used in the scalar QED model. The differences come
from the fact that the target nucleus is no longer treated as point
particles and quarks now carry colors. It is straightforward to see that we
recover the scalar-QED model result $\left( U\left( x_{\perp }\right)
\Rightarrow e^{-igg_{1}\left[ G\left( x_{\perp }\right) \right] }\right) $
if we apply the point particle approximation: $\rho _{a}\left(
z^{-},z_{\perp }\right) t^{a}\Rightarrow \delta \left( z^{-}\right) \delta
^{\left( 2\right) }\left( z_{\perp }\right) $. In addition, the ensemble
average over the color sources should be performed. Since one assumes the
color sources are randomly distributed, a Gaussian distribution $W\left[
\rho \right] $ is always used in the average. In the McLerran-Venugopalan
Model, the Gaussian distribution is defined as follows~\cite{Iancu:2003xm}
\begin{equation}
W\left[ \rho \right] =\mathit{\exp }\left[ -\int {d}z^{-}{d}^{2}z_{\perp }%
\frac{\rho _{a}\left( z^{-},z_{\perp }\right) \rho _{a}\left( z^{-},z_{\perp
}\right) }{2\mu ^{2}\left( z^{-}\right) }\right] .
\end{equation}%
The variance $\mu ^{2}\left( z^{-}\right) $ of the charge distribution
represents the density of color sources per unit volume. It is assumed that
the average in CGC is performed in a functional integration over $\rho $
accompanied by $W\left[ \rho \right] $. It then follows that
\begin{eqnarray}
\langle \rho _{a}\left( x^{-},x_{\perp }\right) \rho _{b}\left(
y^{-},y_{\perp }\right) \rangle _{\rho } &=&\int \mathcal{D}\rho W\left[
\rho \right] \rho _{a}\left( x^{-},x_{\perp }\right) \rho _{b}\left(
y^{-},y_{\perp }\right) \\
&=&\delta _{ab}\delta \left( x^{-}-y^{-}\right) \delta ^{\left( 2\right)
}\left( x_{\perp }-y_{\perp }\right) \mu ^{2}\left( x^{-}\right) .
\end{eqnarray}%
This turns out to be useful in the following derivations for the TMD parton
distributions in CGC.

\subsection{DIS and Drell-Yan Processes}

As we discuss in Sec.II, the TMD quark distribution in DIS in the scalar QED
model can be written as
\begin{eqnarray}
\tilde{q}^{\text{DIS}}\left( x,q_{\perp }\right) &=&\frac{xP^{+2}}{8\pi ^{4}}%
\int dp^{-}p^{-}\int d^{2}R_{\perp }d^{2}R_{\perp }^{\prime }d^{2}r_{\perp
}d^{2}r_{\perp }^{\prime }  \notag \\
&&\times \delta ^{\left( 2\right) }\left( R_{\perp }+r_{\perp }-R_{\perp
}^{\prime }-r_{\perp }^{\prime }\right) e^{iq_{\perp }\cdot \left( R_{\perp
}-R_{\perp }^{\prime }\right) }V\left( r_{\perp }\right) V\left( r_{\perp
}^{\prime }\right)  \notag \\
&&\times \left\{ 1-e^{igg_{1}\left[ G\left( R_{\perp }^{{}}+r_{\perp
}^{{}}\right) -G\left( R_{\perp }\right) \right] }\right\} \left\{
1-e^{-igg_{1}\left[ G\left( R_{\perp }^{\prime }+r_{\perp }^{\prime }\right)
-G\left( R_{\perp }^{\prime }\right) \right] }\right\} \ .  \label{e45}
\end{eqnarray}%
To derive the above result, we have assumed the target hadron (nucleus) to
be a point particle. In order to calculate the quark distribution in CGC, we
need to relax the point particle approximation. Following the above
discussions, we first assume that the target hadron has a color charge
distribution $\rho _{a}\left( z^{-},z_{\perp }\right) $ and perform a
replacement $e^{-igg_{1}\left[ G\left( x_{\perp }\right) \right]
}\Rightarrow U\left( x_{\perp }\right) $.

The second step is to average over the color sources $\rho _{a}\left(
z^{-},z_{\perp }\right) $, which appears in the exponents of the Wilson
lines $U\left( x_{\perp }\right) $, with the Gaussian distribution $W\left[
\rho \right] $. Following this procedure, one finds (see Refs.~\cite%
{Gelis:2001da, Blaizot:2004wv, Fukushima:2007dy})
\begin{eqnarray}
e^{igg_{1}\left[ G\left( R_{\perp }+r_{\perp }\right) -G\left( R_{\perp
}\right) \right] } &\Longrightarrow &\text{Tr}\left\langle U^{\dagger
}\left( R_{\perp }+r_{\perp }\right) U\left( R_{\perp }\right) \right\rangle
_{\rho }  \notag \\
&=&N_{c}\exp \left\{ -\mu _{s}^{2}\int \text{d}^{2}z_{\perp }\left[ G\left(
R_{\perp }+r_{\perp }-z_{\perp }\right) -G\left( R_{\perp }-z_{\perp
}\right) \right] ^{2}\right\}  \notag \\
&=&N_{c}\exp \left\{ -\mu _{s}^{2}\int \text{d}^{2}z_{\perp }\left[ G\left(
r_{\perp }+z_{\perp }\right) -G\left( z_{\perp }\right) \right] ^{2}\right\}
\notag \\
&\simeq &N_{c}\exp \left\{ -Q_{s}^{2}r_{\perp }^{2}/4\right\} ,
\end{eqnarray}%
where $N_{c}=3$ is the number of colors, the saturation scale $Q_{s}$ is
defined as $Q_{s}^{2}=\frac{\mu _{s}^{2}}{2\pi }\ln \frac{1}{r_{\perp
}^{2}\lambda ^{2}}$ with $\mu _{s}^{2}=\frac{g^{2}g_{1}^{2}}{2}%
t^{a}t_{a}\int $d$x^{-}\mu ^{2}\left( z^{-}\right)$. In the
evaluation of the above two-point functions $\left\langle
U^{\dagger }\left( R_{\perp }+r_{\perp }\right) U\left( R_{\perp
}\right) \right\rangle _{\rho }$, we have assumed that the
nucleus size is so large that we can shift $R_{\perp} $ in the
transverse integration. The saturation momentum naturally arises
as a result of multiple scatterings between the hard parton and
color charges inside the nucleus.

The next step is to use fermionic quark splitting kernel instead of the
scalar quark splitting kernel. Thus we replace $V\left( r_{\perp }\right) $
in Eq.~(\ref{e45}) by $\frac{1}{2\pi }2K_{1}\left( Mr_{\perp }\right) $
where the factor of $2$ comes from the fact that fermionic quark has two
different helicities. It is straightforward to derive this fermionic quark
splitting kernel as in Ref.~\cite{Mueller:1999wm}. The rest of the calculation
will remain the same since the eikonal propagator for a fermionic quark is
the same as the one for a scalar quark as in Eq.~(\ref{e5}). After changing
the integral variable to $y=2xP^{+}p^{-}$, we can cast the quark
distribution into
\begin{eqnarray}
x\tilde{q}^{\text{DIS}}\left( x,q_{\perp }\right) &=&\frac{N_{c}}{32\pi ^{6}}%
\int dyd^{2}R_{\perp }d^{2}r_{\perp }d^{2}r_{\perp }^{\prime }e^{-iq_{\perp
}\cdot \left( r_{\perp }-r_{\perp }^{\prime }\right) }\bigtriangledown
_{r_{\perp }}K_{0}\left( \sqrt{y}r_{\perp }\right) \cdot \bigtriangledown
_{r_{\perp }^{\prime }}K_{0}\left( \sqrt{y}r_{\perp }^{\prime }\right)
\notag \\
&&\times \left\{ 1+\exp \left[ -\frac{Q_{s}^{2}\left( r_{\perp }-r_{\perp
}^{\prime }\right) ^{2}}{4}\right] -\exp \left[ -\frac{Q_{s}^{2}r_{\perp
}^{2}}{4}\right] -\exp \left[ -\frac{Q_{s}^{2}r_{\perp }^{\prime 2}}{4}%
\right] \right\} \ .
\end{eqnarray}%
The virtuality of the virtual photon $Q^2=2xP^+P^-$ is taken to be
much larger than $Q_s^2$ and $q_{\perp}^2$. Therefore, one can
approximately integrate $y$ from $0$ to $+\infty$. The dominant
contribution comes from the region where $y$ is close to $0$. It
is hard to evaluate above integrals analytically. Nevertheless, we
can study the quark distribution in the large and small $q_{\perp
}^{2}$ limit, which give
\begin{eqnarray}
\left. \frac{\text{d}x\tilde{q}^{\text{{DIS}}}\left( x,q_{\perp }\right) }{%
d^{2}R_{\perp }}\right\vert _{q_{\perp }^{2}\gg Q_{s}^{2}} &=&\frac{N_{c}}{%
12\pi ^{4}}\frac{Q_{s}^{2}}{q_{\perp }^{2}}  \notag \\
\left. \frac{\text{d}x\tilde{q}^{\text{{DIS}}}\left( x,q_{\perp }\right) }{%
d^{2}R_{\perp }}\right\vert _{q_{\perp }^{2}\ll Q_{s}^{2}} &=&\frac{N_{c}}{%
4\pi ^{4}}\ .  \label{qm}
\end{eqnarray}%
These results agree with those derived in the saturation model for the
quark distribution of a large nucleus in DIS (see e.g., Eqs.(27-29) of Ref.~%
\cite{Mueller:1999wm})\footnote{%
We notice that there is a factor of $1/2$ difference between our results and
those obtained in Ref.~\cite{Mueller:1999wm}. This difference comes from the
fact that the quark distribution calculated in Ref.~\cite{Mueller:1999wm} is
in fact the total quark distribution which includes anti-quark distribution
as well.}. Furthermore, we can transform the above results to the momentum
space and define the normalized unintegrated gluon distribution $F(k_{\perp
},Q_{s})$ as
\begin{equation}
F(k_{\perp },Q_{s})=\int \frac{d^{2}r_{\perp }}{(2\pi )^{2}}e^{-ik_{\perp
}\cdot r_{\perp }}\frac{\text{Tr}\langle U\left( R_{\perp }\right)
U^{\dagger }\left( R_{\perp }+r_{\perp }\right) \rangle _{\rho }}{N_{c}}%
\simeq \frac{1}{\pi Q_{s}^{2}}\exp \left( -\frac{k_{\perp }^{2}}{Q_{s}^{2}}%
\right) \ .\label{fg}
\end{equation}%
In arriving at the Gaussian form of $F(k_{\perp },Q_{s})$ in
Eq.~(\ref{fg}), we have neglected the logarithmic dependence of
$r_{\perp}^2$ in the saturation momentum $Q_s$. Thus, one can
write the quark distribution as a convolution of the unintegrated
gluon distribution and the splitting kernel in momentum space,
\begin{eqnarray}
x\tilde{q}^{\text{DIS}}\left( x,q_{\perp }\right) &=&\frac{N_{c}}{4\pi ^{4}}%
\int d^{2}R_{\perp }d^{2}k_{\perp }F(q_{\perp }-k_{\perp },Q_{s})\int \text{d%
}y\left\vert \frac{\vec{q}_{\perp }}{q_{\perp }^{2}+y}-\frac{\vec{k}_{\perp }%
}{k_{\perp }^{2}+y}\right\vert ^{2}  \notag \\
&=&\frac{N_{c}}{4\pi ^{4}}\int d^{2}R_{\perp }d^{2}k_{\perp }F(k_{\perp
},Q_{s})\left[ 1-\frac{q_{\perp }\cdot (q_{\perp }-k_{\perp })}{q_{\perp
}^{2}-(q_{\perp }-k_{\perp })^{2}}\ln \frac{q_{\perp }^{2}}{(q_{\perp
}-k_{\perp })^{2}}\right] ,  \label{ktf}
\end{eqnarray}%
which is consistent with the results obtained in Ref.~\cite{McLerran:1998nk,
Marquet:2009ca}. The unintegrated gluon distribution $F(k_{\perp },Q_{s})$
is usually defined through the scattering amplitude of a dipole with size $%
r_{\perp }$ on the target nucleus~\cite{Braun:2000wr, Kharzeev:2003wz}. This
dipole scattering amplitude is also equivalent to the expectation value of a
Wilson loop with width $r_{\perp }$ and infinite length as we used above.

For comparison, we can also calculate the quark distribution involved in the
Drell-Yan process. Again, we start with scalar-QED model result~\cite%
{Peigne:2002iw},
\begin{eqnarray}
\tilde{q}^{\text{DY}}\left( x,q_{\perp }\right) &\!\!=\!\!&\frac{xP^{+2}}{%
8\pi ^{4}}\int dp^{-}p^{-}\int d^{2}R_{\perp }d^{2}R_{\perp }^{\prime
}d^{2}r_{\perp }d^{2}r_{\perp }^{\prime }  \notag \\
&&\times \delta ^{\left( 2\right) }\left( R_{\perp }+r_{\perp }-R_{\perp
}^{\prime }-r_{\perp }^{\prime }\right) e^{iq_{\perp }\cdot \left( R_{\perp
}-R_{\perp }^{\prime }\right) }V\left( r_{\perp }\right) V\left( r_{\perp
}^{\prime }\right) \nonumber \\
&&\times \left\{ e^{igg_{1}G\left( R_{\perp }^{{}}+r_{\perp }^{{}}\right)
}-e^{igg_{1}G\left( R_{\perp }\right) }\right\} \left\{ e^{-igg_{1}G\left(
R_{\perp }^{\prime }+r_{\perp }^{\prime }\right) }-e^{-igg_{1}G\left(
R_{\perp }^{\prime }\right) }\right\} \ ,
\end{eqnarray}%
Following the same procedures, we find that the quark distribution in the
Drell-Yan process in the CGC formalism reads as
\begin{eqnarray}
x\tilde{q}^{\text{DY}}\left( x,q_{\perp }\right) &=&\frac{N_{c}}{32\pi ^{6}}%
\int \text{d}y\int d^{2}R_{\perp }d^{2}r_{\perp }d^{2}r_{\perp }^{\prime
}e^{-iq_{\perp }\cdot \left( r_{\perp }-r_{\perp }^{\prime }\right)
}\bigtriangledown _{r_{\perp }}K_{0}\left( \sqrt{y}r_{\perp }\right) \cdot
\bigtriangledown _{r_{\perp }^{\prime }}K_{0}\left( \sqrt{y}r_{\perp
}^{\prime }\right)  \notag \\
&&\times \left\{ 1+\exp \left[ -Q_{s}^{2}\left( r_{\perp }-r_{\perp
}^{\prime }\right) ^{2}/4\right] -\exp \left[ -Q_{s}^{2}r_{\perp }^{2}/4%
\right] -\exp \left[ -Q_{s}^{2}r_{\perp }^{\prime 2}/4\right] \right\} .
\end{eqnarray}%
The quark distribution in the Drell-Yan process is the same as that in DIS,
which is consistent with the conclusion in the scalar QED model, as the QCD
factorization predicts.

\subsection{Dijet production in $pA$ Collisions}

Finally, let us consider the TMD quark distribution for a large nucleus
involved in the di-jet production, again, taking the $qq^{\prime
}\rightarrow qq^{\prime }$ channel as an example. In the scalar-QED model,
the TMD quark distribution in this process is shown in Eq.~(\ref{diq}). In
order to extend to the real QCD calculation, we will assume that the color
charge for quark $q^{\prime }$ is the same as the quark $q$ in the sense of
the average over the large nucleus. This means that we will set $g_{2}=g_{1}$
in the scalar-QED result~\footnote{%
This can be checked against the lowest nontrivial order perturbation expansion of the
multi-gluon exchange contributions in the large $N_{c}$ limit. As we
mentioned before, setting $g_{2}=-g_{1}$ will lead to the quark distribution
in the Drell-Yan process.}. Furthermore, we find that the quark distribution
here will naturally involve four-point function. For example, expanding the
phase factor in Eq.~(\ref{diq}) will depend on the four-point function in
the CGC formalism,
\begin{eqnarray}
&&e^{-igg_{1}\left( G(R_{\perp })-G(R_{\perp }^{\prime })\right) }\left\{
1-e^{igg_{1}\left[ G\left( R_{\perp }^{{}}+r_{\perp }^{{}}\right) -G\left(
R_{\perp }\right) \right] }\right\} \left\{ 1-e^{-igg_{1}\left[ G\left(
R_{\perp }^{\prime }+r_{\perp }^{\prime }\right) -G\left( R_{\perp }^{\prime
}\right) \right] }\right\} \Longrightarrow  \notag \\
&&\left\{
\begin{array}{c}
U\left( R_{\perp }\right) U^{\dagger }\left( R_{\perp }^{\prime }\right)
+U\left( R_{\perp }\right) U^{\dagger }\left( R_{\perp }^{\prime }\right)
U\left( R_{\perp }\right) U^{\dagger }\left( R_{\perp }^{\prime }\right) \\
-U\left( R_{\perp }\right) U^{\dagger }\left( R_{\perp }^{\prime }\right)
U\left( R_{\perp }\right) U^{\dagger }\left( R_{\perp }+r_{\perp }\right)
-U\left( R_{\perp }\right) U^{\dagger }\left( R_{\perp }^{\prime }\right)
U\left( R_{\perp }^{\prime }+r_{\perp }^{\prime }\right) U^{\dagger }\left(
R_{\perp }^{\prime }\right)%
\end{array}%
\right\} \ .
\end{eqnarray}%
The appearance of the four point functions signals the difference between
the di-jet production process and DIS, whereas the latter only involves two
point functions. This indicates that parton distributions directly extracted
from DIS are not sufficient to compute and describe the dijet production
processes.

\begin{figure}[tbp]
\begin{center}
\includegraphics[width=12cm]{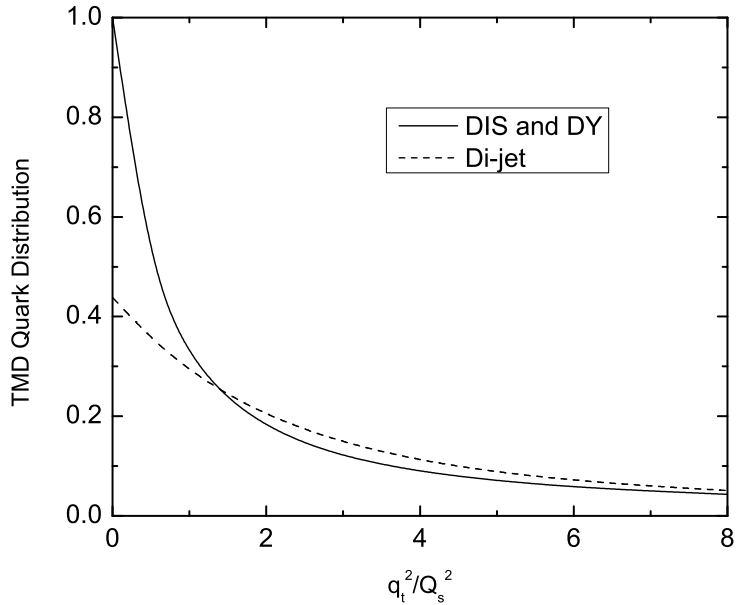}
\end{center}
\caption[*]{Comparison of quark distributions $\frac{4\protect\pi ^{4}}{N_{c}%
}\frac{\text{d}x\tilde{q}\left( x,q_{\perp }\right) }{d^{2}R_{\perp }}$ as
functions of $\frac{q_{\perp }^{2}}{Q_{s}^{2}}$ in DIS (or Drell-Yan) and
di-hadron production. The solid curve stands for the quark distribution in
DIS and Drell-Yan process, and the dash curve represents the distribution
involved in di-hadron production.}
\label{di-dis}
\end{figure}

We can further simplify the above result by taking the large $N_{c}$ limit
for the four point functions~\cite{Dominguez:2008aa}
\begin{equation}
\left\langle U\left( R_{\perp }\right) U^{\dagger }\left( R_{\perp }^{\prime
}\right) U\left( R_{\perp }\right) U^{\dagger }\left( R_{\perp }+r_{\perp
}\right) \right\rangle _{\rho }\simeq \exp \left\{ -\frac{Q_{s}^{2}}{4}\left[
\left( r_{\perp }-r_{\perp }^{\prime }\right) ^{2}+r_{\perp }^{2}\right]
\right\} \ .
\end{equation}%
With this reduction, we arrive at the following quark distribution in the
large $N_{c}$ limit,
\begin{eqnarray}
x\tilde{q}^{\text{DJ}}\left( x,q_{\perp }\right) &=&\frac{N_{c}}{32\pi ^{6}}%
\int dyd^{2}R_{\perp }d^{2}r_{\perp }d^{2}r_{\perp }^{\prime }e^{-iq_{\perp
}\cdot \left( r_{\perp }-r_{\perp }^{\prime }\right) }\bigtriangledown
_{r_{\perp }}K_{0}\left( \sqrt{y}r_{\perp }\right) \cdot \bigtriangledown
_{r_{\perp }^{\prime }}K_{0}\left( \sqrt{y}r_{\perp }^{\prime }\right)
\notag \\
&&\times \left\{
\begin{array}{c}
\exp \left[ -\frac{Q_{s}^{2}\left( r_{\perp }-r_{\perp }^{\prime }\right)
^{2}}{4}\right] +\exp \left[ -\frac{Q_{s}^{2}\left( r_{\perp }-r_{\perp
}^{\prime }\right) ^{2}}{2}\right] \\
-\exp \left[ -\frac{Q_{s}^{2}\left( \left( r_{\perp }-r_{\perp }^{\prime
}\right) ^{2}+r_{\perp }^{2}\right) }{4}\right] -\exp \left[ -\frac{%
Q_{s}^{2}\left( \left( r_{\perp }-r_{\perp }^{\prime }\right) ^{2}+r_{\perp
}^{\prime 2}\right) }{4}\right]%
\end{array}%
\right\} \ ,
\end{eqnarray}%
which then yields%
\begin{eqnarray}
\left. \frac{\text{d}x\tilde{q}^{\text{DJ}}\left( x,q_{\perp }\right) }{%
d^{2}R_{\perp }}\right\vert _{q_{\perp }^{2}\gg Q_{s}^{2}} &=&\frac{N_{c}}{%
12\pi ^{4}}\frac{Q_{s}^{2}}{q_{\perp }^{2}}  \notag \\
\left. \frac{\text{d}x\tilde{q}^{\text{DJ}}\left( x,q_{\perp }\right) }{%
d^{2}R_{\perp }}\right\vert _{q_{\perp }^{2}\ll Q_{s}^{2}} &=&0.44\frac{N_{c}%
}{4\pi ^{4}}\ .
\end{eqnarray}%
It is straightforward to see that the quark distributions in DIS and
di-hadron production have the same perturbative tails while they differ in
the small $q_{\perp }^{2}$ limit. As shown in Fig.~\ref{di-dis}, the quark
distribution is about twice broader than the one in DIS while its peak is
about half of the peak of the DIS distribution. However, it is easy to check
analytically and numerically that the integrated quark distributions are
universal for these processes.

In the momentum space, we find that the quark distribution in di-jet
production can be written as follows:
\begin{eqnarray}
x\tilde{q}^{\text{DJ}}\left( x,q_{\perp }\right) &=&\frac{N_{c}}{4\pi ^{4}}%
\int d^{2}R_{\perp }\int d^{2}l_{\perp }F(q_{\perp }-l_{\perp },Q_{s})
\notag \\
&&\times \int d^{2}k_{\perp }F(k_{\perp },Q_{s})\left[ 1-\frac{l_{\perp
}\cdot (l_{\perp }-k_{\perp })}{l_{\perp }^{2}-(l_{\perp }-k_{\perp })^{2}}%
\ln \frac{l_{\perp }^{2}}{(l_{\perp }-k_{\perp })^{2}}\right] ,  \label{ktf2}
\end{eqnarray}%
which implies that
\begin{equation}
x\tilde{q}^{\text{DJ}}\left( x,q_{\perp }\right) =\int d^{2}l_{\perp }x%
\tilde{q}^{\text{DIS}}\left( x,l_{\perp }\right) F(q_{\perp }-l_{\perp
},Q_{s}).  \label{ktf3}
\end{equation}

This is an interesting new result. It relates the two apparently different
quark distributions through a $k_{t}$ convolution with the unintegrated
gluon distribution $F(q_{\perp }-l_{\perp },Q_{s})$. It is easy to see that
both quark distribution reduce to the same form after integration over $%
q_{\perp }$ since $F(q_{\perp }-l_{\perp },Q_{s})$ is normalized to $1$. In
addition, Eq.~(\ref{ktf3}) explains the broadening of the di-jet quark
distribution as shown in Fig.~\ref{di-dis}. This formula has a natural
physical interpretation. This convolution arises as a result of the extra
initial and final state interactions in the di-jet production process.

\section{Summary and Discussions}

In this paper, we have studied the initial and final state interaction
effects in the small-$x$ parton distributions. As an example, we discussed
the quark distributions in the semi-inclusive deep inelastic scattering,
Drell-Yan lepton pair production and dijet-correlation in $pA$ collisions.
We calculated these distributions first in a scalar-QED model and then
extended to the CGC formalism in QCD.

We have shown the non-universality for the small-$x$ parton distributions
in dijet correlation in the scalar QED model calculations, as compared to the
quark distributions in DIS and Drell-Yan processes.
For the particular partonic channel $qq^{\prime}\rightarrow qq^{\prime }$,
we find that the net effects are summarized into
a phase which leads to a non-vanishing contribution to the quark
distribution and breaks the universality.

We have also calculated the TMD quark distribution involved in dijet production
in the saturation models~\cite{{Iancu:2003xm},Mueller:1999wm}. We reached the
conclusion that TMD quark distributions are not universal in the
color-dipole or color glass condensate formalism, by showing that the quark
distributions involved in DIS and di-jet production processes are distinct
as shown in Fig.~\ref{di-dis}. In addition, we found a simple formula which
relates these two different quark distributions through a convolution with a
normalized gluon distribution.

It is interesting to note that we can also compare to the quark distribution
discussed in Refs.~\cite{mulders}, where the initial and final state
interaction effects are summed into an effective gauge link associated with
the quark distribution. For example, for the partonic channel $%
qq^{\prime}\to qq^{\prime}$, the quark distribution requires the gauge link
as $G=\frac{N_{c}^{2}+1}{N_{c}^{2}-1}\frac{\text{Tr}\left( \mathcal{U}%
^{[\Box ]}\right) }{N_{c}}\mathcal{U}^{[+]}-\frac{2}{N_{c}^{2}-1}\mathcal{U}%
^{[\Box ]}\mathcal{U}^{[-]}$ which is different from that in the
semi-inclusive DIS process with $G=\mathcal{U}^{[+]}$. In the large $N_{c}$
limit, the additional gauge link structure would contribute a factor which
is similar to the unintegrated gluon distribution in the CGC formalism as we
have shown in Sec.III.

Despite the non-universality, we expect that there exists a
generalized TMD factorization for the di-jet production in $pA$ collisions
in the large $A$ limit.
Thanks to the nuclear enhancement, which allows us to neglect any soft gluon
exchanges originated from the proton, we can resum all the anomalous terms
which breaks the $k_{t}$ factorization and put them into the parton
distributions of the target nucleus. This procedure leads to an
effective $k_{t}$ factorization with non-universal nuclear parton
distributions in $pA$ collisions. Also we would like to emphasize that the
Wilson line $U(x_{\perp })$, which provides the underlying fundamental
description of the interactions between partons and dense hadronic matter,
is still universal.

Of course in the high energy limit, the forward di-jet production in $pA$
collisions is dominated by the $qg\rightarrow qg$ channel due to high gluon
density in the target nucleus. This calculation is much more complicated
than the $qq^{\prime }\rightarrow qq^{\prime }$ process which we considered
in this paper. The complexity comes from the fact that there are many more
channels involved in the $qg\rightarrow qg$ process. There have been some
theoretical calculations\cite{Marquet:2007vb,Tuchin:2009nf} in CGC. However,
the non-universality issue has not yet been taken into consideration. We
will address this problem together with the photon-jet productions in $pA$
collisions in a future publication~\cite{Dominguez}.

The non-universality for the TMD parton distributions at small-$x$ clearly
imposes a challenge in explaining the dijet-correlation data in $dA$
collisions at RHIC with the parton distributions extracted \textit{directly}
from the DIS data. The non-universality, on the other hand, provides an
opportunity to study QCD dynamics associated with the initial and final
state interaction effects, which are calculable at small-$x$ (high gluon
density limit) according to our results. More phenomenological discussion
will be provided in ref.~\cite{Dominguez}.

\section{acknowledgment}

We thank Tony Baltz, Stan Brodsky, Fabio Dominguez, Paul Hoyer, Cyrille Marquet,
Larry McLarran,  Al Mueller, Jianwei Qiu and Raju Venugopolan for interesting discussions.
This work was supported in part by the U.S. Department of Energy under
contracts DE-AC02-05CH11231. We are grateful to RIKEN, Brookhaven National
Laboratory and the U.S. Department of Energy (contract number
DE-AC02-98CH10886) for providing the facilities essential for the completion
of this work.

\end{document}